\documentclass{llncs}
\usepackage{llncsdoc}

\usepackage[utf8]{inputenc}
\usepackage{amssymb}
\usepackage{amsmath}
\usepackage{algorithm}
\usepackage[noend]{algpseudocode}
\usepackage{tabularx}
\usepackage{mathtools}
\usepackage{graphicx}
\usepackage{graphicx}
\usepackage[caption=false]{subfig}
\usepackage{hyperref}

\graphicspath{{img/}}

\title{Assessing pattern recognition or labeling in streams of temporal data}
\author{Pierre-François Marteau\inst{1},
}

\institute{IRISA, Université Bretagne Sud, Campus de Tohannic, Vannes, France\\pierre-francois DOT marteau AT univ-ubs DOT fr}
\date{\today}

\begin{document}
\maketitle

\begin{abstract}
In the data deluge context, pattern recognition or labeling in streams is becoming quite an essential and pressing task as data flows inside always bigger streams. The assessment of such tasks is not so easy when dealing with temporal data, namely patterns that have a duration (a beginning and an end time-stamp). This paper details an approach based on an editing distance to first align a sequence of labeled temporal segments with a ground truth sequence, and then, by back-tracing an optimal alignment path, to provide a confusion matrix at the label level. From this confusion matrix, standard evaluation measures can easily be derived as well as other measures such as the "latency" that can be quite important in (early) pattern detection applications.  
\end{abstract}

\section{Introduction}
In the big data and Internet of things era, data is widely generated through streams at an increasing rate. Such data is provided by sensor applications, measured in
network monitoring and traffic management system, available as log records or click-streams in web exploration, sequence of email, blogs data, RSS feeds, social networks, and so many
other sources. Applications requiring pattern recognition in such streams is becoming more and more demanding, and, consequently, the question of assessing such applications is particularly essential. Among various type of stream data we are focusing on stream of temporal data, also called timestamped data stream, for which data elements (samples) are associated with a time index.  The aim of this paper is, given a ground truth sequence of time-stamped labeled segments, to propose a consolidated and extended \textbf{assessment framework} for pattern labeling or recognition in streams. 

\begin{figure}[h!]
  \centering
  \begin{tabular}{cc}
	\includegraphics[width=70mm]{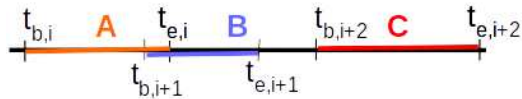}
  \end{tabular}
  \caption{Example of a ground truth sequence.}
\label{fig:GroundTruth}
\end{figure}

We consider in this paper that the ground truth sequence is given as a sequence of time-stamped labels, each label being characterized by a begin and an end time-stamp, as depicted in figure \ref{fig:GroundTruth}. It should be stressed that, to ensure as much as possible the independence of the assessment framework to the nature of the stream, \textbf{such sequence does not keep any reference to the data that are conveyed by the stream} except for the time location of the labeled segments. 

The output of the pattern recognition or labeling tasks in stream that we intend to assess are expected to provide the same kind of sequences. The proposed algorithm will output 
\begin{itemize}
 \item a confusion matrix from which traditional assessment measures such as precision, recall or F-measures can be derived,
 \item an estimation of the recognition latency that can be an issue for real-time application,
 \item an estimation of the average relative duration of the matched labeled segments, that can be of some utility when assessing labeling tasks.
\end{itemize}  

The contribution of this paper is the presentation of a dynamic programming algorithm dedicated to the alignment of such sequences. We insists that it is not one another stream segmentation or labeling method, but a proposal for the assessment of stream segmentation or labeling methods when time occurrence of the labels is a critical issue. The requirement for a ground-truth sequence is a strong constraint that indeed limit the application of the proposed assessment framework, but ensures in general a better acceptance and quality of the assessment since experts are supposedly maintained into the process.

\section{Previous work and problem statement}\label{previous work}

Stream classification or labeling refers essentially to three distinct tasks \cite{Graves2008}\cite{Graves2012}
\begin{itemize}
\item stream classification: the task consists in affecting a class label to a given stream. For instance, categorizing a RSS feed into some pre-defined categories such as sport, economy, society, culture, etc, is typically a stream classification problem.
\item stream segment classification: here we consider that the streams are segmented (the segments are known), and the task consists in affecting a class label to each segment. Considering a general RSS feed, naturally segmented into time-stamped news items, the task is to affect to each news item a class label (e.g. sport, economy, society, etc.).
\item pattern spotting and recognition in streams: here again we consider that the stream is also segmented but the segments are only known for training through manual segmentation and labeling. Hence the segments need to be localized along the time axis, with a beginning and an end time-stamps. This is the case for example when addressing human data such as speech data which can be segmented into phonemes, syllables, words, or any other acoustic or linguistic parts. This tasks is also referred to as temporal classification \cite{Kadous2002}.
\end{itemize}

This paper addresses the assessment of approaches that tackle the third task, obviously the most difficult one since it requires to identify segment frontiers and segment classification as a whole. Formally, such approaches will typically output a sequence of labeled time-stamped segments (SLTSS). 

Let $L$ be the set of labels, and $T$ a time segment. A labeled-segment $s$ will be defined as a triplet $(l,t_b, t_e) \in L \times T \times T$ with the constraint that $t_b \leq t_e$. A sequence of labeled time-stamped segments $S$ will be denoted as \\
$S = s_1 s_2 \cdots s_n = (l_i, t_{b_i}, t_{e_i})_{i=1 \cdots n}$. We will note $S(i)$ the ith labeled segment in the sequence $S$.  

Furthermore, we impose an ordering of the segments inside the sequence, such that: $\forall i$, $t_{b_i} \leq t_{b_{i+1}}$ and $t_{e_i} \leq t_{e_{i+1}}$. From this definition, it should be noted that segments may have different lengths and successive segments can be partially overlapping or disjoint. \\

The problem of assessing pattern spotting and recognition systems in streams comes down to the alignment of a predicted SLTSS with a ground-truth SLTSS, \textbf{with the constraint that only time overlapping segment can be matched}.

If we were dealing uniquely with sequences of symbolic labels, an edit distance, such as the Levensthein \cite{Levenshtein66} or the Smith and Waterman distance \cite{smithWaterman1981} proposed in bioinformatics among others, would be perfectly adapted and has been used for various tasks in bioinformatics, natural language processing, etc, \cite{Graves2012}. However, the time-stamps that delimit the temporal segments introduce a kind of fuzziness in the matching of segments that need to be coped with. It is quite striking that, to our knowledge, no dynamic programming algorithm has been yet proposed to solve this temporal alignment problem at a labeled segment level (straightforwards techniques at frame or sample levels are usually used). Although edit distance based measures have been design to cope with temporal segment sequences such as the time warp edit distance (TWED) \cite{Marteau2009}, these measures does not cope with the label attached to the temporal segments. But more importantly, they are not suited for the alignment of SLTSS we are dealing with, mainly because they enable the alignment of disjoint time segments which has no meaning in our context. The purpose of the temporal alignment algorithm that we detail hereinafter intends to bridge this apparent gap. 

\section{Dynamic programming algorithm for the alignment of a pair of SLTSS}\label{DP-SLTSS}

Following the mathematical definition of the Levenshtein edit distance designed to align two strings we define $\delta_{SLTSS}$ as an edit distance allowing to align two SLTSS $S_1, S_2$ (of length $|S_1|$ and $|S_2|$ respectively) is given by $\operatorname{\delta_{SLTSS}}(S_1(|S_1|),S_2(|S_2|))$ where

\begin{equation}
\label{eq:delta_SLTSS}
\operatorname{\delta_{SLTSS}}(S_1(i),S_2(j)) = \begin{cases}
  C_{0}.\max(i,j) \text{ if} \min(i,j)=0, \\
  \min \begin{cases}
          \operatorname{\delta_{SLTSS}}(S_1(i-1),S_2(j)) + C_{0} \\
          \operatorname{\delta_{SLTSS}}(S_1(i),S_2(j-1)) + C_{0} \\
          \operatorname{\delta_{SLTSS}}(S_1(i-1),S_2(j-1)) + C_{m}(S_1(i), S_2(j))
       \end{cases} \\ \hspace{2in}\text{ otherwise.}\\
\end{cases}\\
\end{equation}

where $\operatorname{\delta_{SLTSS}}(S_1(i),S_2(j))$ is the distance between the first $i$ segments of $S_1$ and the first $j$ segments of $S_2$, $C_{0}>0$ is a constant penalty value  corresponding to a segment insertion or deletion and $C_{m}(S_1(i), S_2(j)) \ge 0$ is the local cost associated to the correct matching of segment $S_1(i)$ with segment $ S_2(j)$ or a substitution of the first segment by the second.\\

Notice that, similarly to the edit distance, but contrary to the DTW measure \cite{Vintsyuk1968}, alignment of one segment with several other (multiple matching of one segment) is not possible. 

\begin{figure}[H]
  \centering
  \begin{tabular}{cc}
	\includegraphics[width=80mm]{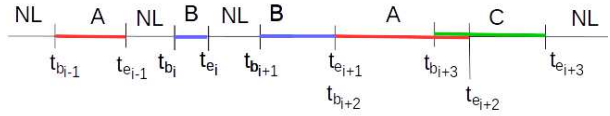}
  \end{tabular}
  \caption{An SLTSS example with contiguous, overlapping and disjoint segments.}
\label{fig:SLTSS}
\end{figure}

The fact that streams can be partially labeled, meaning that labeled segments can be possibly not contiguous (they can be overlapping but also separated by "blank" or "unlabeled" segments), introduces some difficulty. One way to deal with such kind of stream is to introduce a dedicated label NL (for no label) to make the labeling covering the whole time axis. An example of the SLTSS we are working with is given in Figure \ref{fig:SLTSS}. By convention, a NL segment starts at the end of the previous (disjoint) segment and end at the beginning of the following (disjoint) segment. 
 
\begin{figure}[ht!]
  \centering
  \begin{tabular}{cc}
	\includegraphics[width=70mm]{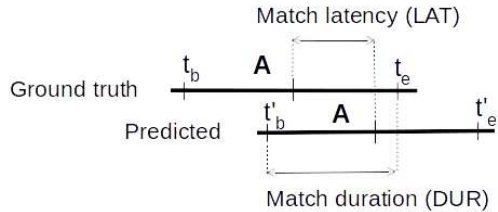}
  \end{tabular}
  \caption{Matching situation involving a ground truth  $A$ labeled segment  aligned with a  predicted $A$ labeled segment.}
\label{fig:matching}
\end{figure}

The main difference with the Levenstein's distance is the local matching cost that is defined for $\delta_{SLTSS}$ as:\\

\begin{equation}
\label{eq:delta_SLTSS}
C_{m}(S_1(i), S_2(j)) = \begin{cases}
\infty \text{ if time intervals } [t_{b_i} ; t_{e_i}] \text{ and } [t_{b_j} ; t_{e_j}] \text{ are disjoint,} \\
\infty \text{ if } l_i \neq l_j \text{ and (} l_i = \text{NL} \text{ or } l_j = \text{NL)}\\
\hspace{1cm} \text{ and } [t_{b_i} ; t_{e_i}] \text{ and } [t_{b_j} ; t_{e_j}] \text{ overlap }\\
\hspace{1cm} \text{  (false positive or false negative),} \\
C_{0} \text{ if } l_i \neq l_j \text{ and } l_i \neq \text{NL} \text{ and } l_j \neq \text{NL}\\
\hspace{1cm} \text{ and } [t_{b_i} ; t_{e_i}] \text{ and } [t_{b_j} ; t_{e_j}] \text{ overlap (substitution),} \\
1.0-\frac{min(t_{e_i}, t_{e_j})-max(t_{b_i}, t_{b_j})}{max(t_{e_i}, t_{e_j})-min(t_{b_i}, t_{b_j})} \text{ otherwise (correct match)} 
\end{cases}\\
\end{equation}

Basically, the local matching cost is evaluated according to the following cases:
\begin{itemize}
\item it is infinite if the segments do not overlap or if the labels are distinct with one being equal to NL (we prohibit the matching of non overlapping segments or NL labeled segments with non NL labeled segments), 
\item it equals $C_0$ if labels are different and distinct to NL, 
\item otherwise it reduces to $1.0$ minus the degree of overlapping of the two segments. Hence, if the two segments are equal (fully overlapping), the cost is null, and if the segments share a single sample, then the cost is $1.0$. \\
\end{itemize}

Figure \ref{fig:matching} shows an example of a partial matching with an overlap that will produce a matching cost verifying $0 < C_m < 1.0$. The $\infty$ cost penalty prohibits the matching of non overlapping segments or segments that are differently labeled. Hence, the substitution operation (switching a label by another) is only allowed if the segments overlap, with a cost of $C_0$.\\

To favor the matching operation over the substitution, insertion and deletion operations, the local penalty cost $C_{0}$ is set up to $2.0$, the double of the matching cost with overlap worse case.\\

 In addition, the average latency estimation (LAT) and the average match duration (DUR) can be estimated. When a match is detected, the latency and the duration measures of the matched segments  are defined accordingly to Figure \ref{fig:matching}. The latency is defined as the  difference of the matched mid-segment time locations, while the match duration is the length of the overlap between the two matched segments.\\

\begin{figure}[h!]
  \centering
  \begin{tabular}{cc}
	\includegraphics[width=70mm]{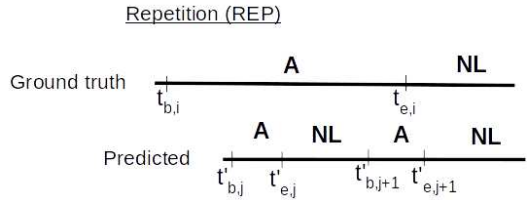}
  \end{tabular}
  \caption{Repetition of a matching label $A$. NL stands for no label.}
\label{fig:repetition}
\end{figure}

\begin{figure}[h!]
  \centering
  \begin{tabular}{cc}
	\includegraphics[width=70mm]{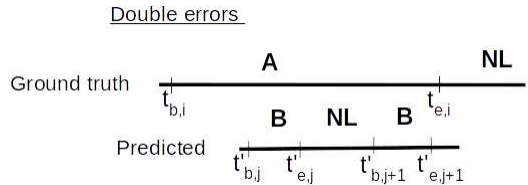}
  \end{tabular}
  \caption{Double errors situation ($A$ v.s. $B B$). NL stands for no label.}
\label{fig:double_errors}
\end{figure}

The evaluation of the errors is performed by \textbf{back-tracing} the best path provided by the recursive equation \ref{eq:delta_SLTSS}. This process is a bit tricky in the presence of repetitions or multiple errors. Repetitions as exemplified in figure \ref{fig:repetition} and possibly multiple errors such as the double errors depicted in figure \ref{fig:double_errors} will produce false positive (FP) and false negative (FN) errors. More precisely one of the repetitions will correspond to a correct match operation while the other occurrences of the repetition will correspond to a deletion (or insertion) operation and will be accounted for FP errors. Similarly, in the presence of multiple errors on a set of overlapping segments, one of the error will be accounted for 1 FN and 1 FP (substitution), while the other errors will be accounted for 1 FP error each. For instance, in figure \ref{DoubleErrors.eps}, the first error will account for 1 FN and 1 FP (substitution A -> B), the second error will account for 1 FP. \\

    \begin{algorithm}[ht!]
        \caption{$\delta_{SLTSS}$: alignment of a ground-truth ($S_1$) and a predicted ($S_2$) labeled segment sequences}\label{DTSS_distance}
        \begin{algorithmic}[1]
            \Procedure{$\delta_{SLTSS}$}{$S_1$, $S_2$} 
               \State \textbf{Double} $D[|S_1|+1][|S_2|+1]$; \Comment{the distance matrix}
               \State \textbf{Integer} $CF[|L|+1][|L|+1]=Zeros(|L|+1,|L|+1)$; \Comment{the confusion matrix}
               \State \textbf{Integer} $match\_count = 0$;
               \For{$i = 0$ to $|S_1|$} $D[i][0] = i\cdot C_{0}$; \Comment{$D$ initialization}
               \EndFor
               \For{$j = 0$ to $|S_2|$} $D[0][j] = j\cdot C_{0}$;
               \EndFor
               \State //$D$ Calculation
               \For{$i = 1$ to $|S_1|$} 
              		\For{$j = 1$ to $|S_2|$} 
              			\State $D[i][j]=min\{D[i-1][j]+C_{0};$
              			\State  \hspace{20mm}   $D[i][j-1]+C_{0};$ 
              			\State  \hspace{20mm}   $D[i-1][j-1]+C_{m}(S_1(i), S_2(j))\};$
               		\EndFor
               \EndFor
               \State //BACK-TRACE
               
               \State $i = |S_1|$, $j = |S_2|;$
               \State $REP=0$, $DUR=0$, $LAT=0;$
			   \While{$i > 0$ \textbf{and} $j > 0$}
			     \State $m = min(D[i-1][j], D[i][j-1], D[i-1][j-1])$;
			     \State \textbf{switch}$(m)$
				\State \hspace{10mm} \textbf{case} $(D[i-1][j-1])$: MATCH/SUBST\_BLOCK; 		
				\State \hspace{10mm} \textbf{break};
				\State \hspace{10mm} \textbf{case} $(D[i-1][j])$: DELETE1\_BLOCK;
				\State \hspace{10mm} \textbf{break};
				\State \hspace{10mm} \textbf{case} $(D[i][j-1])$: DELETE2\_BLOCK;
				\State \hspace{10mm} \textbf{break};
			  \EndWhile
			  \State $DUR=DUR/match\_count$;
			  \State  $LAT=LAT/match\_count$;
			  \State \textbf{return} $CF$, $REP$, $DUR$, $LAT$;
            \EndProcedure
        \end{algorithmic}
    \end{algorithm}

    \begin{algorithm}[h!]
        \caption{MATCH/SUBST\_BLOCK}\label{match_block}
        \begin{algorithmic}[1]
				\If{$S_1(i)$ and $S_2(j)$ have the same label ($l_1(i)=l_2(j)$)} \Comment{correct match}
				    \State $match\_count++$;
				    \State $DUR=DUR+overlap(S_1(i), S_2(j)$;
				    \State $LAT=LAT+latency(S_1(i), S_2(j)$;

%
%
				\EndIf			    
				\State $CF[idx(l_1(i))][idx(l_1(i))]++$; \Comment{$idx(l_1(i))$ is the raw or column index for $l_1(i)$}
				\State $i=i-1$, $j=j-1$;
        \end{algorithmic}
    \end{algorithm}
				
    \begin{algorithm}[h!]
        \caption{DELETE1\_BLOCK}\label{delete1_block}
        \begin{algorithmic}[1]
        	\If{$S_1(j) \neq NL$} \Comment{false negative error}
				\State $CF[0][idx(l_2(j))]++$; 
			\Else \hspace{1mm} NL deletion in GT
			\EndIf
			\State $j=j-1$;
        \end{algorithmic}
    \end{algorithm}

    \begin{algorithm}[h!]
        \caption{DELETE2\_BLOCK}\label{delete2_block}
        \begin{algorithmic}[1]
				\If{$S_2(j) \neq NL$} \Comment{false positive error}
				      \State $CF[idx(l_1(i))][0]++$;
				\Else \hspace{1mm} NL deletion in PRED
				\EndIf
				\State $i=i-1$;
        \end{algorithmic}
    \end{algorithm}
    
The implementation of $\delta_{SLTSS}$ is described in a simplified form in Algorithm \ref{DTSS_distance}, which is decomposed in three blocks depicted in Algorithms \ref{match_block}, \ref{delete1_block} and \ref{delete2_block}\footnote{The code will be made available for the community at the earliest feasible opportunity}. Following are the conventions used to described these algorithms:
\begin{itemize}
    \item the vector or matrix indexes start from 0: basically the first element of a vector $V$ is located at $V[0]$. The same applies for matrix elements.
    \item If $L=\{l_1, l_2, ..., l_{|L|}\}$ is the set of labels (with $NL \notin L$), $l_i$ will correspond to the $i^{th}$ row or column of the confusion matrix (referred to as $CM$ in the algorithms). $idx(l)=i > 0$ will correspond to the index of label $l$ in the $CF$ matrix ($idx(l_i)=i$).
    \item row and column at index $0$ in the $CM$ matrix will correspond to $NL$ (no label). 
\end{itemize}

Algorithm \ref{DTSS_distance} returns a confusion matrix (obtained by back-tracing the best alignment provided by the dynamic programming algorithm), the number of repetitions, the average match duration $DUR$, the average match latency $LAT$. From this confusion matrix it is easy to evaluate assessment measures such as :
\begin{itemize} 
\item Micro Average Accuracy (MAA): $1/|L|\sum_{i=1}^{|L|} (TP_i+TN_i)/(TP_i+FP_i+TN_i+FN_i)$,
\item the Micro Average Precision (MAP): $1/|L|\sum_{i=1}^{|L|} TP_i/(TP_i+FP_i)$,
\item the Micro Average Recall (MAR): $1/|L|\sum_{i=1}^{|L|} TP_i/(TP_i+FN_i)$,
\item Micro Average F1 (MAF1): MAF1 =  2.MAP.MAR/(MAP+MAR).
\end{itemize}

%
%
%

The algorithmic complexity of $\delta_{SLTSS}$ is clearly $O(|S_1|.|S_2|)$, or simply $O(N^2)$ for sequences of length $N$. 

\section{Example}\label{example} 
\begin{figure}[h!]
  \centering
  \begin{tabular}{cc}
	\includegraphics[width=100mm]{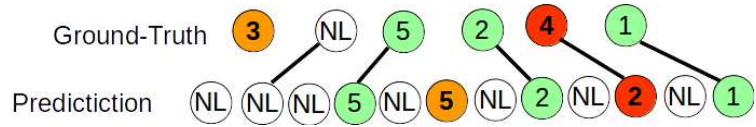}
  \end{tabular}
  \caption{Alignment provided by the algorithm for the pair of SLTSS used in the example. In red/bold the mismatch, in orange/bold the insertion/deletion, in green the correct matches.}
\label{fig:alignment}
\end{figure}

Table \ref{sequences} presents a simple example based on a pair of short sequences. The set of admissible labels is $L=\{1,2,3,4,5\}$. NL is the "no label" extra label. Table \ref{backtrace} present the back-trace process output along a best alignment path. Finally, table \ref{CFM} gives the confusion matrix provided by the backtracking. Clearly, if we except the NL label, three correct matches have been detected, 1 label of the ground truth has not been predicted (label 3), which results in a false negative error, 1 mismatch (substitution) has been detected between labels 2 (GT) and 4 (PRED) and 1 prediction has led to a false positive error (label 5 has been repeated in the predicted sequence). The "best" alignment provided by the algorithm is given in Figure \ref{fig:alignment} 
  
\begin{table}[h!]
\centering
\caption{Ground-truth sequence (left) and predicted sequence (right)}
\label{sequences}
\begin{tabular}{lcr}
\begin{tabular}{|l|l|l|l|}
\hline
index & Label & $t_b$ & $t_e$ \\ \hline
 0 &  3   &  0     &   45     \\ \hline
 1 &  NL   &  46     & 50       \\ \hline
 2 &  5  &    51   &   101     \\ \hline
 3 &  2 &     102  &   152     \\ \hline
 4 &  4  &    153   &  203      \\ \hline
 5 &  1  &    204   &  254      \\ \hline
   
\end{tabular}

& \texttt{           } &

\begin{tabular}{|l|l|l|l|}
\hline
index & Label & $t_b$ & $t_e$ \\ \hline
 0 &   NL   &  0   &  30     \\ \hline
 1 &  NL   &  31  &  50       \\ \hline
 2 &  NL   &  51  &  88     \\ \hline
 3 &  5    &  89  &  90     \\ \hline
 4 &  NL   &  91  &  95      \\ \hline
 5 &  5    &  96  &  106      \\ \hline
 6 &  2    &  107 &  152      \\ \hline
 7 &  NL   &  153 &  174      \\ \hline
 8 &  2    &  175 &  195      \\ \hline
 9 &  NL   &  196 &  203      \\ \hline
 10 &  1    &  204 &  254      \\ \hline
\end{tabular}
\end{tabular}
\end{table}

\begin{table}[h!]
\centering
\caption{\textbf{Back-trace} of the alignment process. PRED means predicted sequence, GT means ground truth sequence. The local alignments are given line by line and each segment is presented as: 
 \textit{sequence\_index} (\textit{label} $|$ $t_b-t_e$)}
\label{backtrace}
\begin{tabular}{|l|l|l|}
\hline
GROUND TRUTH	&	PREDICTION		& EVENTS \\ \hline \hline
5 (1$|$204-254)	&	10 (1$|$204-254)	& correct match \\ \hline
4 (4$|$153-203)	&	9 (NL$|$196-203)	& delete NL in PRED  \\ \hline
4 (4$|$153-203)	&	8 (2$|$175-195)	& mismatch (1 FP and 1 FN)\\ \hline
3 (2$|$102-152)	&	7 (NL$|$153-174)	& delete NL in PRED   \\ \hline
3 (2$|$102-152)	&	6 (2$|$107-152)	& correct match  \\ \hline
2 (5$|$51-101)	&	5 (5$|$96-106)	& correct match  \\ \hline
1 (NL$|$46-50)	&	4 (NL$|$91-95)	& delete NL in PRED  \\ \hline
1 (NL$|$46-50)	&	3 (5$|$89-90)	    & delete in PRED (FP) \\ \hline
1 (NL$|$46-50)	&	2 (NL$|$51-88)	& delete NL in PRED   \\ \hline
1 (NL$|$46-50)	&	1 (NL$|$31-50)	& match NL \\ \hline
0 (3$|$0-45)		&   0 (NL$|$0-30)	    & delete in GT (FN) \\ \hline
0 (3$|$0-45)		&   0 (NL$|$0-30)	    & delete NL in PRED  \\ \hline

\end{tabular}
\end{table}

\begin{table}[h!]
\centering
\caption{\textbf{Confusion matrix} provided by the back-trace of the alignment process. GT stands for ground-truth and PR for predicted}
\label{CFM}
\begin{tabular}{|l|l|l|l|l|l|l|}
\hline
GT/PR & NL&1 & 2 & 3 & 4 & 5 \\ \hline
NL &1 &0 &0 &1 &0 &0  \\ \hline
1&0 &1 &0 &0 &0 &0  \\ \hline
2&0 &0 &1 &0 &1 &0  \\ \hline
3&0 &0 &0 &0 &0 &0  \\ \hline
4&0 &0 &0 &0 &0 &0  \\ \hline
5&1 &0 &0 &0 &0 &1  \\ \hline
\end{tabular}
\end{table}

On this example, the mean latency and mean duration have been respectively evaluated to 9.16 and 35 time units.

\section{Conclusions}\label{conclusions}

The assessment of temporal pattern labeling or recognition tasks in streams is not trivial. It requires to align sequences of labeled time-stamped segments that either partially overlap or are disjoint. We have proposed to use an editing distance to find first an optimal alignment (basically an alignment with a minimal cost) dedicated to this specific alignment problem. Following backward this optimal alignment allows for the inventory of the various kind of mismatch (substitutions, false positive or false negative) errors as well as the correct matches. The choice of the local alignment costs has been made to favor correct label matches and to penalize mismatched labels. 

This algorithm provides a confusion matrix from which standard evaluation measures can be easily derived. Furthermore, it evaluates the average latency and the relative average duration of the correct matches, which can be useful to "rank" applications that require pattern spotting and recognition (e.g. speech or gesture recognition applications) as soon as possible, or a precise temporal location of the labels.

This algorithm requires a ground-truth sequence in input, which is indeed a limitation to its use. In the other hand, since experts are generally involved in the construction of the ground-truth, it ensures a human control of the quality of the assessment procedure and possibly a better acceptance.  



\bibliographystyle{plain}
\bibliography{biblio}

\begin{thebibliography}{1}

\bibitem{Graves2008}
Alex Graves.
\newblock {\em Supervised sequence labelling with recurrent neural networks}.
\newblock PhD thesis, Technical University Munich, 2008.

\bibitem{Graves2012}
Alex Graves.
\newblock {\em Supervised Sequence Labelling with Recurrent Neural Networks},
  volume 385 of {\em Studies in Computational Intelligence}.
\newblock Springer, 2012.

\bibitem{Kadous2002}
Mohammed~Waleed Kadous.
\newblock {\em Temporal Classification: Extending the Classification Paradigm
  to Multivariate Time Series}.
\newblock PhD thesis, New South Wales, Australia, Australia, 2002.
\newblock AAI0806481.

\bibitem{Levenshtein66}
Vladimir~Iosifovich Levenshtein.
\newblock Binary codes capable of correcting deletions, insertions and
  reversals.
\newblock {\em Soviet Physics Doklady}, 10(8):707--710, feb 1966.
\newblock Doklady Akademii Nauk SSSR, V163 No4 845-848 1965.

\bibitem{Marteau2009}
P.~F. Marteau.
\newblock Time warp edit distance with stiffness adjustment for time series
  matching.
\newblock {\em IEEE Transactions on Pattern Analysis and Machine Intelligence},
  31(2):306--318, Feb 2009.

\bibitem{smithWaterman1981}
Temple~F. Smith and Michael~S. Waterman.
\newblock Identification of common molecular subsequences.
\newblock {\em Journal of Molecular Biology}, 147(1):195--197, 1981.

\bibitem{Vintsyuk1968}
T.~K. Vintsyuk.
\newblock Speech discrimination by dynamic programming.
\newblock {\em Cybernetics}, 4(1):52--57, 1968.

\end{thebibliography}

\end{document}